\documentclass[journal]{IEEEtranTIE}
\usepackage{graphicx}
\usepackage{cite}
\usepackage{picinpar}
\usepackage{amsmath}
\usepackage{url}
\usepackage{flushend}
\usepackage[utf8]{inputenc}
\usepackage{colortbl}
\usepackage{soul}
\usepackage{multirow}
\usepackage{pifont}
\usepackage{color}
\usepackage{alltt}
\usepackage[hidelinks]{hyperref}
\usepackage{enumerate}
\usepackage{siunitx}
\usepackage{breakurl}
\usepackage{epstopdf}
\usepackage{pbox}
\usepackage{caption}
\usepackage{algorithm}
\usepackage{algpseudocode}

\begin{document}

\title{Low-cost Pyranometer-Based ANN Approach for MPPT in Solar PV Systems}

\author{
	\vskip 1em
	Luiz Fernando M. Arruda, \emph{Student Membership},
	Moisés Ferber, \emph{Membership},
	\\ and Diego Greff, \emph{Membership} \\
	\vskip 1.8em
	\large{\textbf{Post Conference Paper}}
}

\maketitle


\begin{abstract}
    This article presents a study on the application of artificial neural networks (ANNs) for maximum power point tracking (MPPT) in photovoltaic (PV) systems using low-cost pyranometer sensors. The proposed approach integrates pyranometers, temperature sensors, and an ANN to estimate the duty cycle of a DC/DC converter, enabling the system to consistently operate at its maximum power point. The strategy was implemented in the local control of a \'Cuk converter and experimentally validated against the conventional Perturb and Observe (P\&O) method. Results demonstrate that the ANN-based technique, leveraging affordable sensor technology, achieves accurate MPPT performance with reduced fluctuations, enhancing the responsiveness and efficiency of PV tracking systems.
\end{abstract}
\section{Introduction}\label{sec:intro}

Solar energy is one of the fastest-growing sectors in the global energy landscape, having experienced a remarkable 1135.15\% increase from 2013 to 2023. Despite this growth, integrating photovoltaic (PV) systems into the electrical grid remains challenging. Key issues include weather dependence, power conversion losses, grid instability, and the complexity of grid synchronization~\cite{10542138}. Additional challenges involve maximum power point tracking (MPPT) under dynamic conditions and partial shading caused by clouds, dust~\cite{10537168}, and physical obstructions such as buildings and trees~\cite{10586857}.

To address these challenges, a variety of MPPT algorithms have been developed to improve the efficiency of PV modules under rapidly changing weather and irradiance conditions. These techniques can be classified into classical, hybrid, optimal, and intelligent methods~\cite{9171659}. Classical approaches, such as Constant Voltage (CV), Incremental Conductance (IC), Open-Circuit Voltage (OCV), Short-Circuit Current (SCC), Hill-Climbing (HC), Perturb and Observe (P\&O), Modified P\&O, and Adaptive Reference Voltage (ARV), are relatively simple to implement in embedded systems and rely on voltage and/or current measurements~\cite{9237072}.

The application of artificial intelligence (AI) in solving electrical engineering problems has become increasingly common, particularly due to advances in embedded computing and microcontroller technologies. Notable examples include adaptive parameter optimization in STATCOMs using ANNs~\cite{8086222}, data-driven energy management in plug-in hybrid electric vehicles~\cite{8717709}, fusion-based dynamic modeling of proton exchange membrane fuel cells~\cite{9851451}, and AI-assisted control strategies for enhancing transient response in power-electronic-dominated grids~\cite{10275316, 10475519}.

Several studies have investigated AI-based MPPT techniques~\cite{8651555, 7926354, 8258969, 7917311, 6975159}, each proposing distinct methodologies to estimate the optimal operating point. In this context, the present work introduces an ANN-based MPPT strategy that uses temperature and solar irradiance measurements from a low-cost sensor to predict the ideal duty cycle of a power converter.

The main contributions of this work are summarized as follows:
\begin{itemize}
    \item The use of a low-cost pyranometer for real-time solar irradiance estimation;
    \item The application of artificial neural networks to directly estimate the duty cycle in power electronic converters, eliminating the need for perturbative methods;
    \item An experimental comparison between the conventional P\&O method and the proposed ANN-based MPPT technique using low-cost sensors.
\end{itemize}

To validate the proposed methodology, it was implemented in C and embedded into a \'Cuk DC/DC converter. Experimental tests were conducted to evaluate and compare its performance.

The remainder of this paper is organized as follows. Section~\ref{sec:review} presents an overview of artificial neural networks (ANNs). Section~\ref{sec:app} describes the implementation of the proposed methodology. Section~\ref{sec:results} discusses the experimental results. Finally, Section~\ref{sec:concl} presents the conclusions.
\section{ANN Methodology\label{sec:review}}

The core concept of the proposed methodology relies on using a low-cost pyranometer sensor combined with an artificial neural network (ANN). ANNs act as universal function approximators and are highly effective in solving complex nonlinear problems.

\begin{figure}[!ht]
    \centering
    \includegraphics[scale=1]{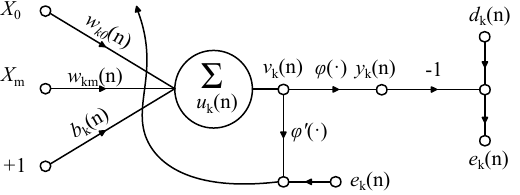}
    \caption{{\fontsize{9}{9}\selectfont Artificial Neural Model}}
    \label{fig:artificialneuralmodel}
\end{figure}

A model of an artificial neuron is shown in Figure~\ref{fig:artificialneuralmodel}. In this study, the artificial neuron employs the generalized delta rule for weight adjustment~\cite{haykin2004comprehensive, arruda2020aplicaccao, 9684052, da2021controlador}, which assumes that the error is directly measurable, thus enabling each synaptic weight to be updated individually.

Similar to a biological neuron, $X_m$ represents the input stimuli, $w_{km}$ the synaptic weights, and $d_k(n)$ the desired output. The synaptic weight determines the importance of each input signal in the processing. The activation value of the artificial neuron $u_k$ is calculated as:

\begin{equation}
    u_k(n) = \sum_{j=0}^{m} w_{kj}(n) \cdot x_j(n)
\end{equation}

A bias value $b_k$ is then added to modify the activation potential:

\begin{equation}
    v_k(n) = u_k(n) + b_k
\end{equation}

The neuron output $y_k$ is obtained by applying a nonlinear activation function $\varphi(\cdot)$:

\begin{equation}
    y_k(n) = \varphi(v_k(n))
\end{equation}

This activation function limits the output to a finite range. In this work, the hyperbolic tangent function is used:

\begin{equation}
    \varphi(v_k(n)) = \frac{e^{v_k} - e^{-v_k}}{e^{v_k} + e^{-v_k}}
\end{equation}

The training of an ANN involves adjusting the network weights based on a dataset composed of known input-output pairs. The objective is to make the output $y_k$ as close as possible to the desired output $d_k$~\cite{haykin2004comprehensive}.

The forward propagation of inputs through the network layers to the output is known as the feed-forward process. For each training instance, the error $e_k$ is computed as:

\begin{equation}
    e_k(n) = d_k(n) - y_k(n)
\end{equation}

The mean squared error (MSE), used as a performance metric, is given by:

\begin{equation}
    \varepsilon(n) = \frac{1}{2} e_k^2(n)
\end{equation}

The learning process occurs via backpropagation, where the error is propagated backward through the network layers to update the synaptic weights. The local gradient $\delta(n)$ for each weight is calculated by:

\begin{equation}
    \delta(n) = e_k(n) \cdot \varphi'_k(v_k(n))
\end{equation}

Since the activation function is the hyperbolic tangent, its derivative is:

\begin{equation}
    \varphi'_k(v_k(n)) = 1 - \tanh^2(v_k(n))
\end{equation}

Using the generalized delta rule, each synaptic weight is updated as:

\begin{equation}
    w_{kj}(n+1) = w_{kj}(n) + \alpha \cdot \delta(n) \cdot x_j(n)
\end{equation}

where $\alpha$ is the learning rate. 

To ensure generalization, the dataset is divided into two subsets: a training set (typically 70\% of the data) and a validation set (the remaining 30\%). The selection of the neural network architecture in this work was performed through trial-and-error analysis. Various architectures were evaluated using different activation functions, including GELU, ReLU, SELU, sigmoid, softmax, softplus, softsign, and swish. The configuration with the best generalization capability or lowest mean squared error was selected for implementation.

\begin{figure}[!ht]
    \centering
    \includegraphics[width=0.5\textwidth]{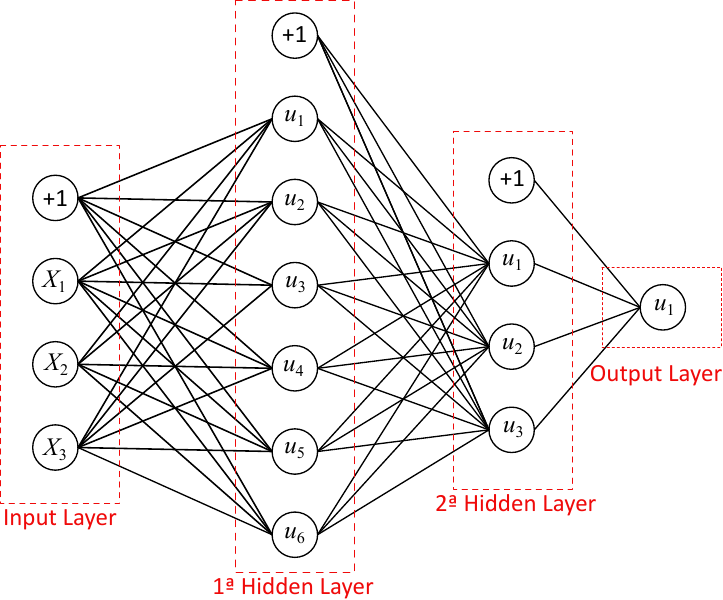}
    \caption{{\fontsize{9}{9}\selectfont Multilayer Perceptron}}
    \label{fig:redeneuralartificial}
\end{figure}

Figure~\ref{fig:redeneuralartificial} illustrates a multilayer perceptron (MLP), which consists of multiple layers of neurons. Each neuron processes and propagates its output, along with the bias, to the neurons in the next layer. This process continues layer by layer until the final (output) layer is reached. An MLP structure was adopted in this work.

\begin{algorithm}
    \caption{Train and Validate ANN} 
    \label{alg:fluxoTreinoRNA}
    \begin{algorithmic}[1]
        \State \textbf{Start}
        \State \textbf{Create Dataset} (Solar Irradiation, Temperature, Duty Cycle)
        \State \textbf{Split Dataset} into Training and Validation sets
        \State $i \gets 0$
        \While{$i < 5$}
            \State \textbf{Choose} an ANN architecture and activation function
            \State \textbf{Train} ANN with the training dataset
            \State \textbf{Validate} ANN with the validation dataset
            \State \textbf{Record} ANN characteristics and performance
            \State $i \gets i + 1$
        \EndWhile
        \State \textbf{Select} the best-performing ANN
        \State \textbf{Export} the weights of the selected ANN
        \State \textbf{Implement} the ANN in C on the control board
        \State \textbf{End}
    \end{algorithmic}
\end{algorithm}

Algorithm~\ref{alg:fluxoTreinoRNA} summarizes the training and validation procedure. It outlines a systematic approach that begins with dataset preparation and splitting. The process involves iterative testing of different ANN configurations to identify the most accurate model. The selected ANN's weights are then exported and embedded in the control system, ensuring practical deployment in a microcontroller environment.
\section{Application of the Methodology in a Power Electronic Converter\label{sec:app}}

This section is divided into three parts: the low-cost pyranometer description, dataset creation, and neural network training.

\subsection{Low-cost Pyranometer}

\begin{figure}[!htb]
    \centering
    \includegraphics[scale=.4]{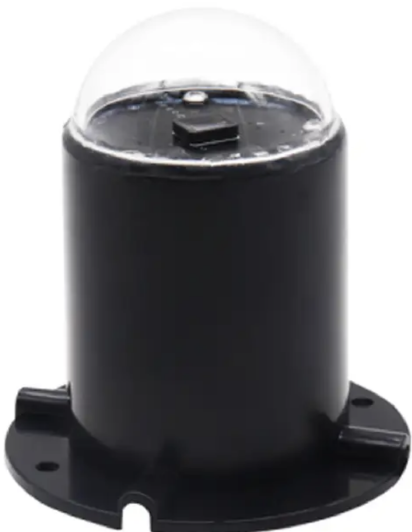}
    \caption{Low-cost solar irradiance sensor RS-RA-V05-JT}
    \label{fig:solar_irradiance_sensor}
\end{figure}

The low-cost sensor used in this study is the RA-T5-V05-JT, shown in Fig.~\ref{fig:solar_irradiance_sensor}. It converts solar irradiance into an output voltage signal and, when used with the ANN, represents a key innovation of this work. By using geographical parameters (latitude, longitude, timezone, and day of the year) and physical data (solar irradiance, Sun radius, Earth-Sun distance, air mass, ozone layer absorption, and direct irradiance), it is possible to estimate solar irradiance throughout the day. This estimation enables proper sensor calibration~\cite{pveducation2019}.

\begin{table}[!htb]
    \centering
    \caption{Geographic and Physical Data for Estimating Solar Irradiance}
    \label{table:solar_irradiation}
    \begin{tabular}{|l|c|}
        \hline
        Parameter & Value \\ \hline
        Latitude & -26.2348783 \\
        Longitude & -48.886931 \\
        Timezone & -3 \\
        Day of the Year & 91 (April 4\textsuperscript{th}) \\
        Solar Irradiance & $62.3 \times 10^6$ W/m$^2$ \\
        Sun Radius & $695 \times 10^6$ m \\
        Earth–Sun Distance & $149.5 \times 10^{9}$ m \\
        Air Mass & 1.16 \\
        Absorption and Dispersion & 30\% \\
        Direct Irradiance & 70\% \\ \hline
    \end{tabular}
\end{table}

\begin{figure}[!ht]
    \centering
    \includegraphics[scale=1]{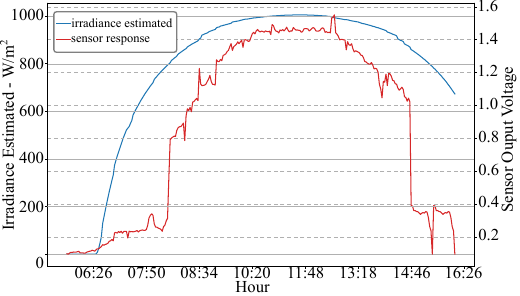}
    \caption{Comparison between Estimated Irradiance and Sensor Output}
    \label{fig:solar_irradiance_sensor_Extimated}
\end{figure}

Using these parameters and a data logger, Fig.~\ref{fig:solar_irradiance_sensor_Extimated} shows the estimated solar irradiance (in W/m\(^2\)) and the sensor output voltage throughout the day. The x-axis represents the time (from 06:00 to 16:00), the left y-axis shows estimated irradiance, and the right y-axis shows the sensor voltage. The blue curve indicates the estimated irradiance, following a typical diurnal solar pattern: starting at zero before sunrise, peaking near noon, and decreasing toward sunset. The red curve represents the sensor response, which generally follows the same pattern but with noticeable fluctuations, possibly due to shading, cloud cover, or sensor noise. Between 08:30 and 15:30, partial shading from nearby buildings significantly reduces the sensor output.

\subsection{Dataset Creation}

To obtain reliable data that captures system behavior under various conditions, a data logger is typically used in a photovoltaic (PV) system to monitor key variables. The main environmental inputs affecting the PV system operating point are solar irradiance and temperature. Combined with the construction data of the PV panel, it is possible to estimate voltage, current, and power at the maximum power point (MPP).

\begin{table}[!ht]
    \centering
    \caption{YL150P-17B Panel Characteristics}
    \begin{tabular}{|l|c|c|}
        \hline
        Parameter & Symbol & Value \\ \hline
        Maximum Power & $P_{max}$ & 150 W \\
        Efficiency & $\eta$ & 15\% \\
        Voltage at $P_{max}$ & $V_{mpp}$ & 18.5 V \\
        Current at $P_{max}$ & $I_{mpp}$ & 8.12 A \\
        Open-Circuit Voltage & $V_{oc}$ & 22.9 V \\
        Short-Circuit Current & $I_{sc}$ & 8.61 A \\
        Temperature Coefficient of $V_{oc}$ & $\beta$ & -0.37\%/°C \\
        Temperature Coefficient of $I_{sc}$ & $\alpha$ & 0.06\%/°C \\
        Number of Cells & – & 36 \\ \hline
    \end{tabular}
    \label{tab:panel_YL150P-17B}
\end{table}

In this work, mathematical equations representing the behavior of the YL150P-17B panel (see Table~\ref{tab:panel_YL150P-17B}) were used. Temperature and irradiance data were collected and used to calibrate the pyranometer and generate MPP voltage, current, and power values.

Since the ultimate goal is to predict the duty cycle of a power converter, it was also necessary to include static gain and load parameters. The dataset was constructed using estimated voltage and power values, varying resistive loads (1~$\Omega$ to 19~$\Omega$ in steps of 2~$\Omega$), and calculating the corresponding duty cycle for a \'Cuk converter. The dataset spans irradiance values from 100 to 1000~W/m\(^2\) and temperatures from 5.11~°C to 60.93~°C, resulting in 797,700 total records.

The final dataset contains four variables: solar irradiance, temperature, load resistance (inputs), and the converter duty cycle (output).

\begin{figure}[!ht]
    \centering
    \includegraphics[scale=1]{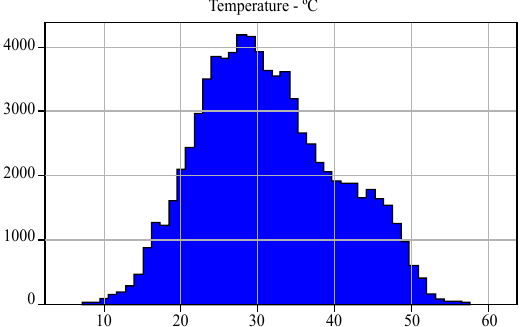}
    \caption{Temperature Histogram}
    \label{fig:conjuntoTreinamento_Temperature}
\end{figure}

\begin{figure}[!ht]
    \centering
    \includegraphics[scale=1]{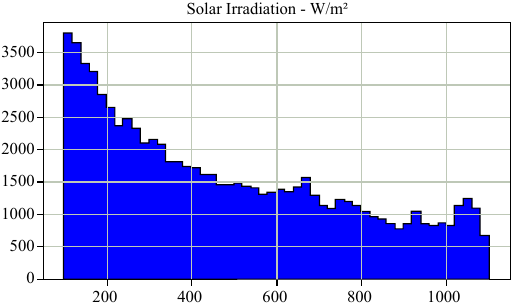}
    \caption{Solar Irradiation Histogram}
    \label{fig:conjuntoTreinamento_SolarIrradiation}
\end{figure}

\begin{figure}[!ht]
    \centering
    \includegraphics[scale=1]{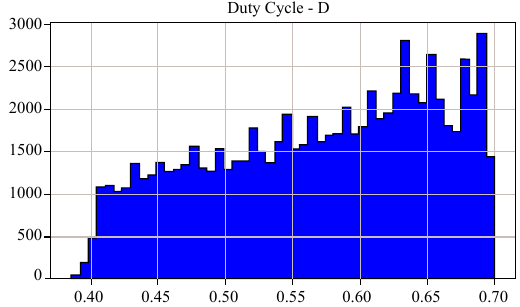}
    \caption{Duty Cycle Histogram}
    \label{fig:conjuntoTreinamento_DutyCycle}
\end{figure}

\begin{figure}[!ht]
    \centering
    \includegraphics[scale=1]{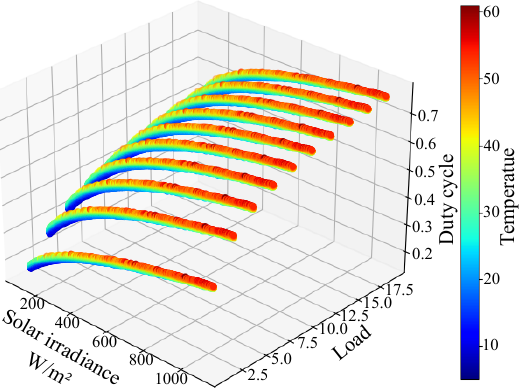}
    \caption{Correlation Between Dataset Variables}
    \label{fig:curva_duty_temo_inc_with_r}
\end{figure}

Figures~\ref{fig:conjuntoTreinamento_Temperature} to~\ref{fig:curva_duty_temo_inc_with_r} show the distribution and interaction of these parameters. Temperature readings (Fig.~\ref{fig:conjuntoTreinamento_Temperature}) are mostly centered around 29°C, with fewer extreme values. Fig.~\ref{fig:conjuntoTreinamento_SolarIrradiation} shows a high concentration of low-irradiance data, typical of early morning and late afternoon. Fig.~\ref{fig:conjuntoTreinamento_DutyCycle} reveals that most duty cycles fall between 0.60 and 0.68.

Fig.~\ref{fig:curva_duty_temo_inc_with_r} presents the 3D correlation among the variables. It can be observed that as irradiance increases, so does the temperature. The duty cycle is highly sensitive to irradiance and load variation. Under high irradiance (near 1000~W/m\(^2\)), the duty cycle stabilizes around 0.6–0.7. At low irradiance (around 200~W/m\(^2\)), duty cycles are more widely spread (0.2–0.6). Similarly, lower resistive loads (e.g., 2~$\Omega$) exhibit greater duty cycle variation, reinforcing the need for dynamic adjustment in MPPT strategies.

\subsection{Artificial Neural Network}

The ANN was developed and trained using the Python programming language and TensorFlow~\cite{tensorflow2015-whitepaper}. Each training process lasted between 3 and 5 hours over 14 epochs. The dataset was divided into 70\% training and 30\% validation.

The selected ANN architecture consists of three input neurons (irradiance, temperature, resistance), two hidden layers with 6 and 3 neurons respectively, and one output neuron representing the duty cycle. The hyperbolic tangent function was used as the activation function.

Multiple activation functions were tested. Linear activation yielded a high mean squared error (MSE) of 4.50\%, while GELU reduced the error to 0.13\%. Other functions (ReLU, SELU, sigmoid, softmax, softplus, softsign, and swish) produced MSEs ranging from 0.05\% to 0.86\%.

\begin{figure}[!ht]
    \centering
    \includegraphics[scale=1]{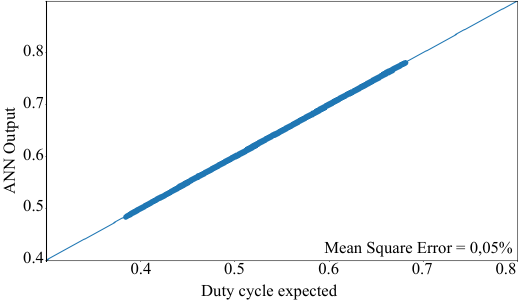}
    \caption{ANN Output vs. Validation Dataset (Normalized)}
    \label{fig:tensorflow_tanh_norm}
\end{figure}

Fig.~\ref{fig:tensorflow_tanh_norm} shows the network output compared to the validation data (normalized). The ideal prediction lies along the diagonal. Points closely aligned with the diagonal indicate high predictive accuracy.

\begin{figure}[!ht]
    \centering
    \includegraphics[scale=1]{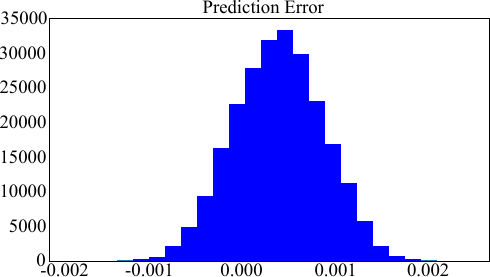}
    \caption{Prediction Error Distribution}
    \label{fig:tensorflow_tanh_norm_error}
\end{figure}

After training, the prediction errors were analyzed. Fig.~\ref{fig:tensorflow_tanh_norm_error} shows the frequency of outputs with absolute error less than 0.002, confirming the model’s precision.

In conclusion, the choice of activation function and network architecture significantly affects ANN performance. The hyperbolic tangent function, used in the final model, demonstrated excellent generalization and low prediction error, validating its effectiveness for MPPT applications in PV systems.
\section{Experimental Results\label{sec:results}}

\begin{figure}[!ht]
    \centering
    \includegraphics[scale=1]{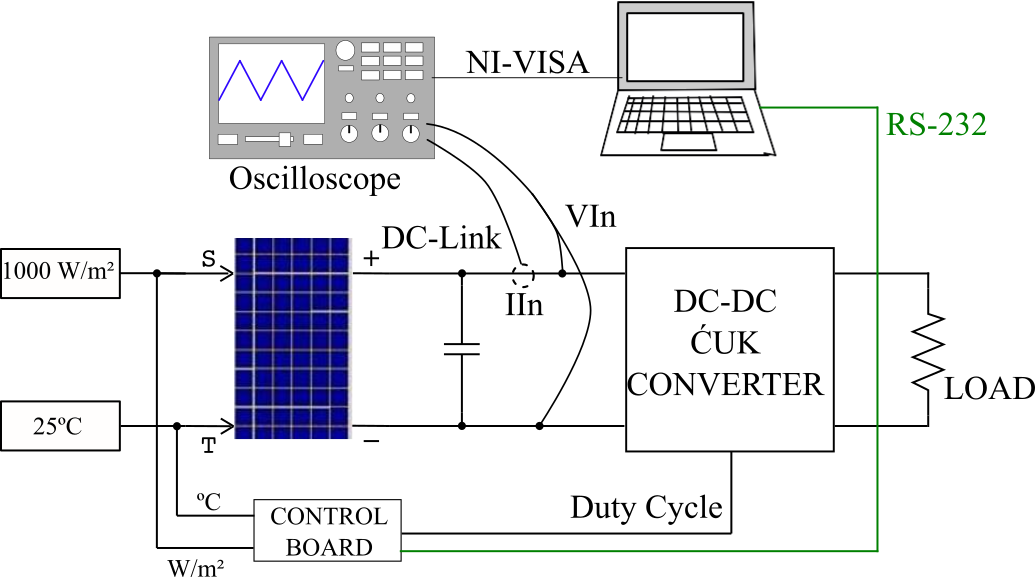}     
    \caption{Overview of ANN-based MPPT data acquisition and analysis}
    \label{fig:Fig_mppt_RNA} 
\end{figure}

\begin{figure}[!ht]
    \centering
    \includegraphics[scale=1]{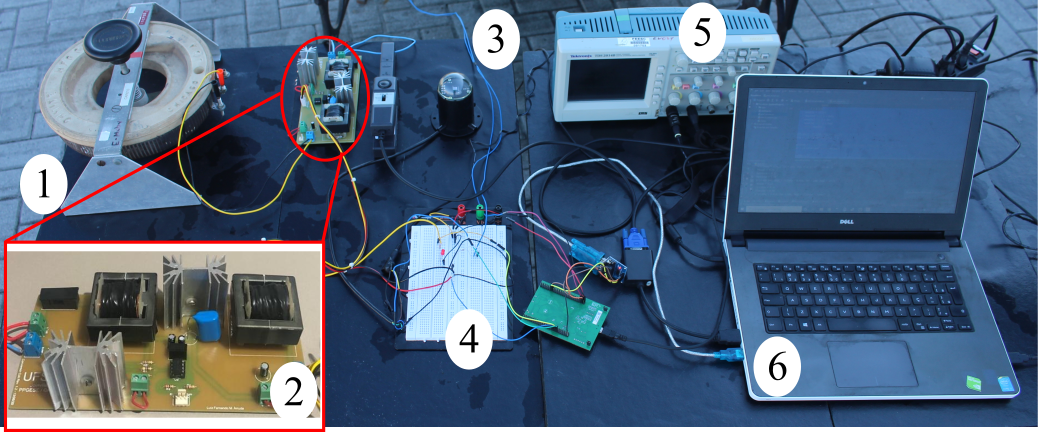}
    \caption{Experimental setup: (1) Resistive load, (2) \'Cuk DC-DC converter, (3) Solar irradiance sensor, (4) STM32F407VG with serial communication, (5) Oscilloscope, (6) Computer}
    \label{fig:foto_bancada_aquisicao}
\end{figure}

Figures~\ref{fig:Fig_mppt_RNA} and~\ref{fig:foto_bancada_aquisicao} show the experimental setup used to validate the proposed methodology. The pyranometer and temperature sensors are directly connected to the STM32F407VG control board, which in turn controls the power electronic converter. Voltage and current sensors are connected to a Tektronix oscilloscope for waveform monitoring. The oscilloscope interfaces with a computer via NI-VISA for experiment logging. Additionally, irradiance and temperature data are transmitted from the control board to a laptop over a serial RS-232 connection, completing the data acquisition process.

\begin{figure}[ht!]
    \centering
    \includegraphics[scale=1]{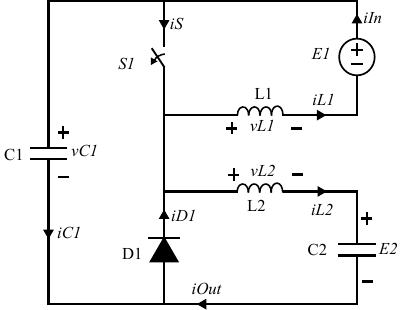}
    \caption{{\fontsize{9}{9}\selectfont \'Cuk Converter Circuit}}
    \label{fig:cukConverter}
\end{figure}

The converter used to validate the ANN-based MPPT strategy is a \'Cuk converter, illustrated in Fig.~\ref{fig:cukConverter}. The design and component parameters are summarized in Tables~\ref{tab:projetoCuk} and~\ref{tab:componentesCuk}, respectively.

\begin{figure}[ht!]
    \centering
    \begin{minipage}[b]{0.22\textwidth}
        \centering
        \small
        \captionof{table}{{\fontsize{9}{9}\selectfont Parameters}}
        \label{tab:projetoCuk}
        \begin{tabular}{|c|c|}
            \hline
            Parameter        &  Value     \\ \hline
            $E_1$            & 18.5~V     \\ 
            $P_S$            & 150~W      \\ 
            $P_O$            & 150~W      \\ 
            $E_2$            & -38.729~V  \\ 
            $f_{sw}$         & 100~kHz    \\ \hline
        \end{tabular}
    \end{minipage}
    \hfill
    \begin{minipage}[b]{0.22\textwidth}
        \centering
        \small
        \captionof{table}{{\fontsize{9}{9}\selectfont Components}}
        \label{tab:componentesCuk}
        \begin{tabular}{|c|c|}
            \hline
            Component & Value       \\ \hline
            $L_1$     & 308~$\mu$H  \\ 
            $L_2$     & 3232~$\mu$H \\ 
            $C_1$     & 5~$\mu$F    \\ 
            $C_2$     & 25~nF       \\ \hline
        \end{tabular}
    \end{minipage}
\end{figure}

\begin{figure}[!ht]
    \centering
    \includegraphics[scale=1]{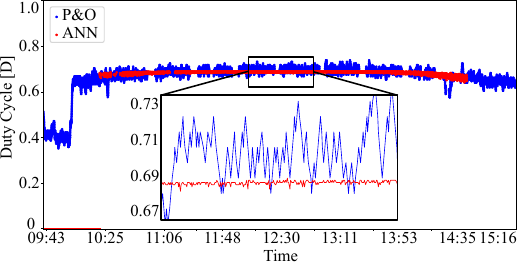}
    \caption{Duty cycle comparison between P\&O and ANN-based MPPT methods}
    \label{fig:EXP_Duty_Cycle_zoom} 
\end{figure}

Figure~\ref{fig:EXP_Duty_Cycle_zoom} compares the duty cycle behavior between the classical P\&O method and the proposed ANN-based strategy. The Perturb and Observe (P\&O) algorithm operates by periodically disturbing the operating point and measuring the resulting change in output power, inherently producing oscillations around the MPP. Conversely, the ANN method directly estimates the optimal duty cycle based on irradiance, temperature, and load conditions, yielding a much smoother and more consistent profile throughout the day.

\begin{figure}[!ht]
    \centering
    \includegraphics[scale=1]{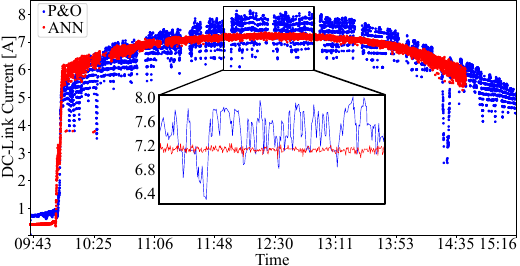}
    \caption{DC-Link current comparison: P\&O vs. ANN}
    \label{fig:EXP_Corrente_Entrada_Total_zoom}
\end{figure}

Figure~\ref{fig:EXP_Corrente_Entrada_Total_zoom} presents the DC-Link current profile. Although overall efficiency was not the primary focus, it is evident that the ANN-based method results in a more stable current waveform, reducing ripple and electrical stress on components—thereby potentially extending the system's operational life.

\begin{figure}[!ht]
    \centering
    \includegraphics[scale=1]{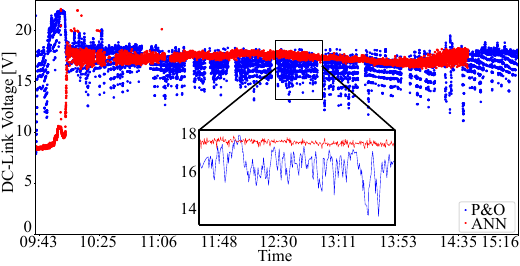}
    \caption{DC-Link voltage comparison between P\&O and ANN strategies}
    \label{fig:EXP_Tensao_Total_zoom}
\end{figure}

\begin{figure}[!ht]
    \centering
    \includegraphics[scale=1]{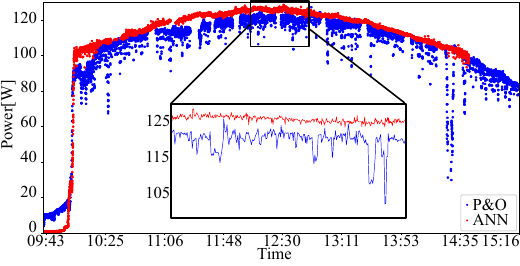}
    \caption{Power output comparison between classical P\&O and ANN-based MPPT}
    \label{fig:EXP_Potencia_Total_zoom}
\end{figure}

Figure~\ref{fig:EXP_Tensao_Total_zoom} shows the DC-Link voltage response. The ANN maintains consistent voltage levels compared to P\&O, contributing to greater system stability. Figure~\ref{fig:EXP_Potencia_Total_zoom} displays the power output throughout the same period, reinforcing that the ANN technique maintains superior and more stable power delivery under varying irradiance conditions.

The joint analysis of duty cycle, current, voltage, and power clearly highlights the superior dynamic performance of the ANN-based MPPT technique. The integration of low-cost irradiance sensors with ANN prediction improves tracking precision, responsiveness, and overall converter efficiency, even under highly variable environmental conditions.
\section{Conclusion\label{sec:concl}}

The application of Artificial Neural Networks (ANNs) in power electronics—particularly for Maximum Power Point Tracking (MPPT) in photovoltaic systems—demonstrates significant advantages, especially when combined with low-cost sensor technology. The experimental results confirm that ANNs can effectively track the maximum power point with minimal fluctuations in duty cycle, DC-Link current, output power, and voltage. By leveraging historical data on irradiance, temperature, and power, ANN-based controllers can accurately predict optimal operating conditions, outperforming traditional MPPT techniques.

The integration of low-cost irradiance sensors further enhances the practicality of the proposed approach, offering a reliable and efficient solution for real-time energy optimization. Compared to the widely used Perturb and Observe (P\&O) algorithm, the ANN-based method exhibits superior stability and performance, maintaining higher and more consistent power output while minimizing dynamic oscillations.

Moreover, the observed voltage and current stability suggest reduced stress on system components, which may extend their operational lifespan and lower maintenance requirements. The combination of advanced machine learning algorithms and affordable hardware presents a scalable, cost-effective alternative for MPPT implementation—suitable for a wide range of applications, including off-grid systems and decentralized renewable energy sources. This approach contributes to the broader goal of enhancing energy efficiency and promoting sustainable energy practices.


\bibliographystyle{Bibliography/IEEEtranTIE}
\bibliography{references}
\section{Biography Section}

\vspace{-33pt}

\begin{IEEEbiography}
[{\includegraphics[width=1in,height=1.25in,clip,keepaspectratio]{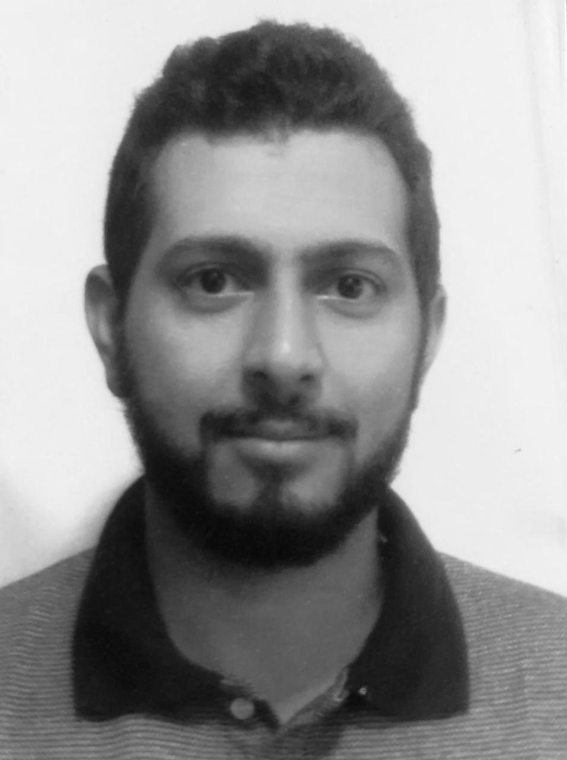}}]
{Luiz Fernando Marquez Arruda}
born in Goiânia, Brazil, in 1989. He received a B.Eng. degree in Computer Engineering by the Centro Universitário de Goiás (UNI-ANHANGUERA), Goiânia, Goiás in 2012 and the Master's degree in Electronic Systems by Federal University of Santa Catarina - UFSC, in Joinville, Brazil in the year 2021.

He is currently a Ph.D. Student in Electrical and Computer Engineering at University of Chicago Illinois at Chicago, Chicago, IL, United States. His research interests are power electronics applied to renewable energy and cybersecurity.
\end{IEEEbiography}

\vspace{-33pt}

\begin{IEEEbiography}
[{\includegraphics[width=1in,height=1.25in,clip,keepaspectratio]{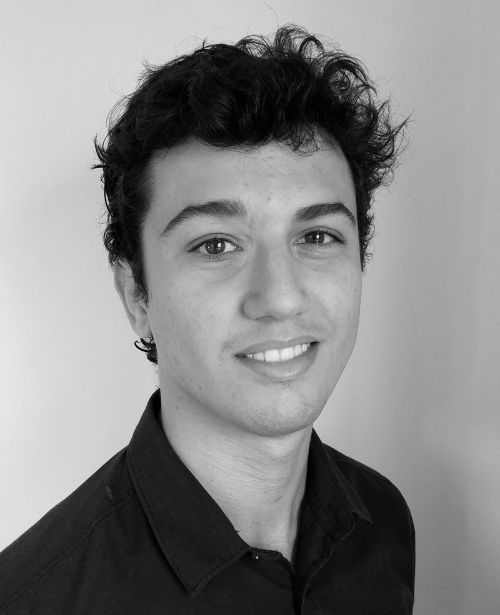}}]
{Moisés Ferber}
was born in Belo Horizonte, Brazil, in 1985. He received the B.Eng. and Doctoral degrees in electrical engineering from the Federal University of Minas Gerais, Belo Horizonte, Brazil, in 2010 and 2013, respectively. He completed the Doctoral degree in a joint supervision with the Laboratoire Ampere, Ecole Centrale de Lyon, Ecully, France.

He is currently an Assistant Professor with the Federal University of Santa Catarina, Joinville, Brazil. His research interests include uncertainty quantification, electromagnetic compatibility, and renewable energy.
\end{IEEEbiography}

\vspace{-33pt}

\begin{IEEEbiography}
[{\includegraphics[width=1in,height=1.25in,clip,keepaspectratio]{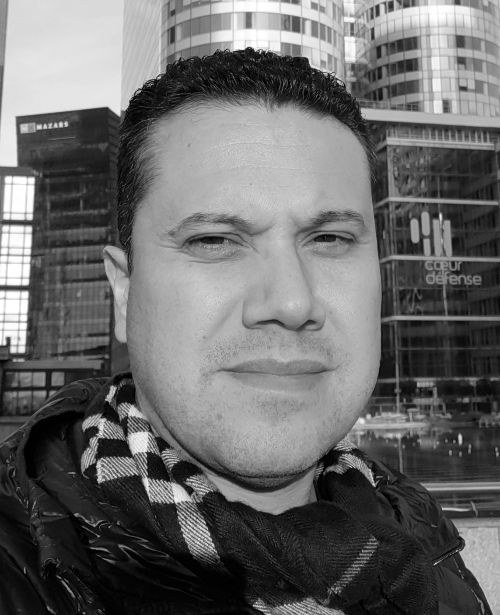}}]
{Diego Greff}
Graduated and Masters in Electrical Engineering from the Federal University of Santa Maria (1997), and Ph.D. in Electrical Engineering from the Federal University of Santa Catarina (2009). 

He worked for 11 years in the multinational private sector, Epcos, Renault, Whirlpool, developing products, managing global and national projects, and managing people. He is currently an adjunct professor at the Federal University of Santa Catarina at the Joinville Campus and a member of the Research Group on Drives and Control of Mechatronic Systems. He has experience in Electrical Engineering, with an emphasis on Industrial Electronics, working mainly on the following topics: static converters, three-phase PWM rectifiers, power factor correction, research, development, and execution of embedded electronics in consumer goods.
\end{IEEEbiography}

\vfill

\end{document}